\documentclass[11pt]{article}
\usepackage{amssymb,latexsym,amsmath}
\usepackage{amsfonts, graphics}
\newcommand{\R}{\mathbb R}

\def\open#1{\setbox0=\hbox{$#1$}
\baselineskip = 0pt
\vbox{\hbox{\hspace*{0.4 \wd0}\tiny $\circ$}\hbox{$#1$}}
\baselineskip = 11pt\!}
\def\fn{\open{f}}

\def\be{\begin{equation}}
\def\ee{\end{equation}}
\def\bea{\begin{eqnarray}}
\def\eea{\end{eqnarray}}
\def\beas{\begin{eqnarray*}}
\def\eeas{\end{eqnarray*}}

\def\supp{\mathrm{supp}\,}

\newtheorem{theorem}{Theorem}
\newtheorem{lemma}{Lemma}

\newtheorem{corollary}{Corollary}

\def\open#1{\setbox0=\hbox{$#1$}
\baselineskip = 0pt
\vbox{\hbox{\hspace*{0.4 \wd0}\tiny $\circ$}\hbox{$#1$}}
\baselineskip = 11pt\!}
\def\fn{\open{f}}
\def\supp{\mathrm{supp}\,}

\begin{document}
\title{Black hole formation from a complete regular past for collisionless matter}
\author{H\aa kan Andr\'{e}asson\\
Mathematical Sciences\\ University of Gothenburg\\
        Mathematical Sciences\\ Chalmers University of Technology\\
        S-41296 G\"oteborg, Sweden\\
        email: hand@chalmers.se}
\maketitle

\begin{abstract}
Initial data for the spherically symmetric Einstein-Vlasov system is constructed whose past evolution is
regular and whose future evolution contains a black hole. This is the first example of initial data
with these properties for the Einstein-matter system with a "realistic" matter model. One consequence of the result is that there exists
a class of initial data for which the ratio of the Hawking mass $\open{m}=\open{m}(r)$ and the area
radius $r$ is arbitrarily small everywhere, such that a black hole forms in the evolution. This result is in a sense analogous to
the result~\cite{Cu3} for a scalar field. Another consequence is that there exist black hole initial data such that the solutions
exist for all Schwarzschild time $t\in (-\infty,\infty)$.
\end{abstract}

\section{Introduction}
\setcounter{equation}{0}
An important question in the study of gravitational collapse is to identify physically admissible initial data,
and it is natural to require that the past evolution of the data is regular. However, most of the existing mathematical results
which ensure a regular past also ensure a regular future, cf.~\cite{RR,CK}, which rules out the study of the formation of black holes. The exceptions being the classical 
result for dust~\cite{OS}, where some classes of solutions have a regular 
past, and the recent result~\cite{Ds2} for a scalar field. In the latter work, which 
in part
rests on the studies~\cite{Cu2,Cu3}, initial data whose past evolution is
regular and whose future evolution forms a black hole is constructed.
Now, neither dust nor a scalar field are realistic matter models in the sense that they are
used by astrophysicists. Dust is a perfect fluid where the pressure is assumed to be zero, and a scalar field is merely
a toy model. Thus, there is so far no example of a solution to the Einstein-matter system for a realistic matter model
possessing a regular past and a singular future.

In this work we consider collisionless matter governed by the Vlasov equation, cf.~\cite{A1} and~\cite{R1} for an introduction.
Although this is a simple matter model, it has
rich dynamics and many features that are desirable of a realistic matter model. Indeed, it allows for anisotropic pressure, there is a large number of stable and unstable spherically symmetric and axially symmetric stationary solutions~\cite{AKR4,Rn1}, there is numerical support that time periodic solutions exist~\cite{AR2}, it behaves as Type I matter in critical collapse~\cite{AR2,OC,RRS2}, and it is used by astrophysicists~\cite{BT}. The following theorem is the main result of the present paper. 
\begin{theorem}\label{Theorem1}
There exists a class of initial data ${\cal J}$ for the spherically symmetric Einstein-Vlasov system with the property that
black holes form in the future time direction and in the past time direction spacetime is causally geodesically
complete.
\end{theorem}
We will see that a
consequence of this result is that for any $\epsilon>0$, initial data can be constructed with the property that the ratio $\open{m}/r$ of
the initial Hawking mass $\open{m}=\open{m}(r)$, and the area radius $r$, is less than $\epsilon$ everywhere, such that a black hole forms in
the evolution. We formulate this as a corollary.
\begin{corollary}\label{Corollary1}
Given $\epsilon>0$, there exists a class ${\cal J}_r$ of initial data for the spherically symmetric Einstein-Vlasov system which
satisfy
\[
\sup_r \frac{\open{m}\,(r)}{r}\leq \epsilon,
\]
for which black holes form in the evolution.
\end{corollary}
This result improves the main result of~\cite{AKR2} and is analogous to the result~\cite{Cu3} in the case of a scalar field where conditions
on the data are given which ensure the formation of black holes. These conditions give no lower bound on $2\open{m}/r$ but involve other restrictions.
Another consequence of our result is the following corollary.
\begin{corollary}\label{Corollary2}
There exists a class ${\cal J}_s$ of black hole initial data for the spherically symmetric Einstein-Vlasov system such that the corresponding solutions exist for all Schwarzschild time $t\in (-\infty,\infty)$.
\end{corollary}
In the future time direction this corollary was shown in~\cite{AKR2}, the improvement here is that the solutions exist on the entire real line.

The present result relies in part on the previous studies~\cite{A2},~\cite{AKR1} and~\cite{AKR2}, which now will be reviewed. In~\cite{AKR1} global existence
in a maximal time gauge is shown for a particular class of initial data where the particles are moving rapidly outwards.
One of the restrictions imposed on the initial data is that
\begin{equation}\label{ratiomtg}
\sup_{r}\frac{2\open{m}(r)}{r}<k_0,
\end{equation}
where the constant $k_0$ is roughly $1/10.$
The situation considered in~\cite{AKR2} is in a sense the reverse since the initial data is such that the particles move rapidly inwards
and the quantity $\sup_{r} 2\open{m}/r$ is required to be close to one. The main result of~\cite{AKR2} is that data of this kind guarantee the formation
of black holes in the evolution. The analysis in~\cite{AKR2} is carried out in Schwarzschild coordinates, i.e. in a polar time gauge. Now,
particles that move inward in the future time direction move outward in the past time direction. It is thus natural to try to combine these two results
with the goal of constructing solutions with a regular past and a singular future. The conditions on the ratio $2\open{m}/r$ are clearly very
different in~\cite{AKR1} compared to~\cite{AKR2}, and moreover, the Cauchy hypersurfaces are different since a maximal
time gauge and a polar time gauge are imposed in the respective cases. The main reason why a maximal time gauge is used in~\cite{AKR1} is
due to the difficulties related to the so called pointwise terms in the characteristic equations in Schwarzschild coordinates. In~\cite{A2} the problem of global
existence for general initial data is investigated under conditional assumptions on the solutions. The analysis
along characteristics is applied to a modified quantity for which the problems with the pointwise terms in Schwarzschild coordinates do not appear.

In the present
work we combine the strategies in~\cite{A2} and~\cite{AKR1} and show global existence for rapidly outgoing particles in Schwarzschild coordinates.
In particular the result in~\cite{AKR1} is improved by showing that the restriction (\ref{ratiomtg}) can be relaxed, and for
sufficiently fast moving particles $2\open{m}/r$ is allowed to be arbitrarily close to one. By combining this result with the result in~\cite{AKR2} we are then able
to construct data whose past is regular and whose future contains a black hole.

The outline of the paper is as follows. In the next section the spherically symmetric Einstein-Vlasov system is introduced. Global existence for rapidly outgoing 
particles is shown in section 3 for two different sets of initial data which are 
adapted to Corollary~1 and Corollary~2 respectively. Finally, in section 4 the proofs of Theorem~\ref{Theorem1}, Corollary~1 and Corollary~2 are given.

\section{The Einstein-Vlasov system}
\setcounter{equation}{0}
For an introduction to the Einstein-Vlasov system and kinetic theory we refer to~\cite{A1} and~\cite{R1}, and for a careful derivation of the system given below we refer to~\cite{R}.
In Schwarzschild coordinates the spherically symmetric metric takes the form
\begin{equation}
ds^{2}=-e^{2\mu(t,r)}dt^{2}+e^{2\lambda(t,r)}dr^{2}
+r^{2}(d\theta^{2}+\sin^{2}{\theta}d\varphi^{2}).
\end{equation}
The Einstein equations read
\begin{eqnarray}
&\displaystyle e^{-2\lambda}(2r\lambda_{r}-1)+1=8\pi r^2\rho,&\label{ee1}\\
&\displaystyle e^{-2\lambda}(2r\mu_{r}+1)-1=8\pi r^2 p,&\label{ee2}\\
&\displaystyle\lambda_{t}=-4\pi re^{\lambda+\mu}j,&\label{ee3}\\
&\displaystyle e^{-2\lambda}(\mu_{rr}+(\mu_{r}-\lambda_{r})(\mu_{r}+
\frac{1}{r}))-e^{-2\mu}(\lambda_{tt}+\lambda_{t}(\lambda_{t}-\mu_{t}))=
8\pi p_T.&\label{ee4}
\end{eqnarray}
The indices $t$ and $r$ denote partial derivatives.
The Vlasov equation for the density function
$f=f(t,r,w,L)$ is given by
\begin{equation}
\partial_{t}f+e^{\mu-\lambda}\frac{w}{E}\partial_{r}f
-(\lambda_{t}w+e^{\mu-\lambda}\mu_{r}E-
e^{\mu-\lambda}\frac{L}{r^3E})\partial_{w}f=0,\label{vlasov}
\end{equation}
where
\begin{equation}
E=E(r,w,L)=\sqrt{1+w^{2}+L/r^{2}}.\label{E}
\end{equation}
Here $w\in (-\infty,\infty)$ can be thought of as the radial component
of the momentum variables, and $L\in [0,\infty)$ is the square of
the angular momentum.
The matter quantities are defined by
\begin{eqnarray}
\rho(t,r)&=&\frac{\pi}{r^{2}}
\int_{-\infty}^{\infty}\int_{0}^{\infty}Ef(t,r,w,L)\;dwdL,\label{rho}\\
p(t,r)&=&\frac{\pi}{r^{2}}\int_{-\infty}^{\infty}\int_{0}^{\infty}
\frac{w^{2}}{E}f(t,r,w,L)\;d
wdL,\label{p}\\
j(t,r)&=&\frac{\pi}{r^{2}}
\int_{-\infty}^{\infty}\int_{0}^{\infty}wf(t,r,w,L)\;dwdL,\label{j}\\
p_T(t,r)&=&\frac{\pi}{2r^{4}}\int_{-\infty}^{\infty}\int_{0}^{\infty}\frac{L}{E}f(t,r,w,L)\;
dwdL.\label{q}
\end{eqnarray}
Here $\rho,p,j$ and $p_T$ are the energy density, the radial pressure, the current and the tangential pressure respectively.
The following boundary conditions are imposed to ensure asymptotic flatness
\begin{equation}
\lim_{r\rightarrow\infty}\lambda(t,r)=\lim_{r\rightarrow\infty}\mu(t,r)=0,\label{bdryaf}
\end{equation}
and a regular centre requires that
\begin{equation}\label{regcentre}
\lambda(t,0)=0.
\end{equation}
We point out that the Einstein equations are not independent and that
e.g. the equations (\ref{ee3}) and (\ref{ee4}) follow by
(\ref{ee1}), (\ref{ee2}) and (\ref{vlasov}).

As initial data it is sufficient to prescribe a density function
$\open{f}=\open{f}(r,w,L)\geq 0$ such that
\begin{equation}\label{notsinit}
   \int_0^r 4\pi\eta^2\open{\rho}\,(\eta)\,d\eta < \frac{r}{2}.
\end{equation}
Here we denote by $\open{\rho}$ the energy density induced by the initial
distribution function $\open{f}$. This condition ensures that no trapped surfaces 
are present initially. Given $\open{f},$ equations (\ref{ee1}) and (\ref{ee2}) can be solved to give $\lambda$ and $\mu$ at $t=0.$
We will only consider initial data
such that $\open{f}(r,\cdot,L)=0$ if $r\leq\epsilon$, for some $\epsilon>0$, or if $L\geq L_+,$ for some $L_+>0.$ If the initial data is $C^1([\epsilon,\infty[,]-\infty,\infty[,[0,\infty[)$ we say that it is regular. 

Let us write down a couple of
facts about the system (\ref{ee1})-(\ref{bdryaf}).
A solution to the Vlasov
equation can be written
\begin{equation}
f(t,r,w,L)=\open{f}(R(0,t,r,w,L),W(0,t,r,w,L),L),
\label{repre}
\end{equation}
where $R$ and $W$ are solutions of the characteristic system
\begin{eqnarray}
\frac{dR}{ds}&=&e^{(\mu-\lambda)(s,R)}\frac{W}{E(R,W,L)},\label{char1}\\
\frac{dW}{ds}&=&-\lambda_{t}(s,R)W-e^{(\mu-\lambda)(s,R)}\mu_{r}(s,R)E(R,W,L)
\nonumber\\
& &+e^{(\mu-\lambda)(s,R)}\frac{L}{R^3E(R,W,L)},\label{char2}
\end{eqnarray}
such that $(R(s,t,r,w,L),W(s,t,r,w,L),L)=(r,w,L)$ when $s=t$.
This representation shows that $f$ is nonnegative for all $t\geq 0,$ $\|f\|_{\infty}=\|\open{f}\|_{\infty},$ and that $f(t,r,w,L)=0$ if $L>L_{+}.$
The Hawking mass $m$ of the sphere of area radius $r$ is given by 
\begin{equation}
m(t,r)=4\pi\int_{0}^{r}\eta^{2}\rho(t,\eta)d\eta,\label{m}
\end{equation}
and by integrating (\ref{ee1}) we find
\begin{equation}
e^{-2\lambda(t,r)}=1-\frac{2m(t,r)}{r}.\label{e2-lambda}
\end{equation}
A fact that we will need is that $$\mu+\lambda\leq 0.$$
This is easily obtained by adding the equations (\ref{ee1}) and
(\ref{ee2}), which gives $$\lambda_{r}+\mu_{r}\geq 0,$$ and then using
the boundary conditions (\ref{bdryaf}). Furthermore, from (\ref{e2-lambda})
we get that $\lambda\geq 0,$ and it follows that $\mu\leq 0.$ We also introduce the notations
$\hat{\mu}$ and $\check{\mu}.$ From equation (\ref{ee2}) and (\ref{bdryaf}) we have
\begin{equation}\label{muhatcheck}
\mu(t,r)=-\int_r^{\infty}\frac{m(t,\eta)}{\eta^2}e^{2\lambda}\,d\eta-\int_r^{\infty}4\pi \eta pe^{2\lambda}\, d\eta=:\hat{\mu}+\check{\mu}.
\end{equation}
We will need an expression for $\hat{\mu}_t$. By (\ref{ee3}) and (\ref{e2-lambda}) it follows that $m_t(t,r)=-4\pi r^2 j(t,r)e^{\mu-\lambda}$, and we obtain 
\begin{equation}\label{dtmuhat}
\hat{\mu}_t(t,r)=\int_r^{\infty}4\pi j(t,\eta) e^{(\mu+\lambda)(t,\eta)}e^{2\lambda(t,\eta)}d\eta.
\end{equation}
An important quantity is the ADM mass $M$, given by 
\begin{equation}
M=4\pi\int_{0}^{\infty}r^{2}\rho(t,r)dr.\label{adm}
\end{equation}
The fact that it is conserved follows by using (\ref{ee3}) and (\ref{e2-lambda}). 

The following result is given in \cite{A2} but since the proof is short we
include it for completeness. By a regular solution we mean a solution which
is launched by regular initial data with compact support.
\begin{lemma}
Let $(f,\mu,\lambda)$ be a regular solution to the Einstein-Vlasov system.
Then
\begin{eqnarray}
&\displaystyle\int_{0}^{\infty}4\pi r(\rho+p)e^{2\lambda}e^{\mu+\lambda}dr\leq 1,&\label{expr}\\
&\displaystyle\int_{0}^{\infty}(\frac{m}{r^2}+4\pi rp)e^{2\lambda}e^{\mu}dr\leq
1.&\label{expr2}
\end{eqnarray}
\end{lemma}
\textbf{Proof. }Using the boundary condition (\ref{bdryaf}) we get
\begin{eqnarray*}
1\geq
1-e^{\mu+\lambda}(t,0)&=&\int_{0}^{\infty}\frac{d}{dr}e^{\mu+\lambda}dr\\
&=&\int_{0}^{\infty}(\mu_{r}+\lambda_{r})e^{\mu+\lambda}dr.
\end{eqnarray*}
The right hand side equals (\ref{expr}) by equations (\ref{ee1})
and (\ref{ee2}) which completes the first part of the lemma. The
second part follows by considering $e^{\mu}$ instead of
$e^{\mu+\lambda}.$
\begin{flushright}
$\Box$
\end{flushright}
Finally, we note that in~\cite{RR} and~\cite{A2} local existence theorems are proved for compactly supported 
and non-compactly supported initial data respectively, 
and it will be used below that solutions exist on some time interval $[0,T[.$
\section{Global existence for outgoing matter}
In order to understand the mechanism behind the global existence result
for outgoing matter we recall the example in~\cite{AKR1} and consider the much simpler
Vlasov-Poisson system which is the Newtonian limit of the
Einstein-Vlasov system. Due to the spherical symmetry the
maximal force experienced by a particle at distance $r$
from the origin is $-M/r^2$ in the Vlasov-Poisson case,
where $M>0$ is the total mass of the particle ensemble.
Hence along any particle trajectory
\[
\frac{d}{dt} \left(\frac{1}{2} w^2 - \frac{M}{r}\right) =
 w \dot w + \frac{M}{r^2}\dot r = w\,\left(\dot w + \frac{M}{r^2}\right)
\geq  0,
\]
as long as its radial velocity $\dot r = w= x\cdot v/r \geq 0$.
Hence
\[
\frac{1}{2} w^2(t) - \frac{M}{r(t)} \geq \frac{1}{2} w^2(0) - \frac{M}{r(0)}
\]
and
\[
\frac{1}{2} w^2(t) \geq \frac{1}{2} w^2(0) - \frac{M}{r(0)}
\]
on any time interval on which $w(t)$ remains non-negative.
Now let $w_1>0$ be an initial lower bound for the radial velocities
of the particles in the ensemble, $r_1>0$ an initial lower bound for
their distance from the origin,
and assume that
\[
W_1 := \frac{1}{2} w_1^2 - \frac{M}{r_1} > 0.
\]
Then as long as a particle is moving outward,
\[
w(t) > W_1,\ r(t) > r_1 + W_1 t.
\]
This implies that all the particles keep moving outward for
all future time.

Let us turn back to the spherically symmetric Einstein-Vlasov system.
The notation below is adapted to the notation in~\cite{AKR2} since the aim is to 
show that the initial data
we construct overlap with the initial data in~\cite{AKR2}. 
Two different sets of initial data, adapted to Corollary 1 and 2, will be considered 
and two similar results on global existence will be shown below; 
Theorem 2 and Theorem 3. 
Let us point out that if the only goal had been to improve the global existence 
result for rapidly outgoing particles in~\cite{AKR1} then we could have considered 
a simpler class of initial data analogous to the data in~\cite{AKR1}, 
cf. Remark 2 below. 

Let $0<r_0<r_1$ be given and put $M=r_1/2$. Let $\open{f}_s$ be data of 
a steady state supported in $[r'_0,r_0]$ and let 
\begin{equation}\label{M-checkM}
M_\mathrm{in}:=\int_{r'_0}^{r_0}4\pi r^2\open{\rho}_s(r)dr, 
\end{equation}
where $\open{\rho}_s$ is induced by $\open{f}_s$. 
The results in~\cite{A4} guarantee that such steady states exist if $r'_0$ 
is sufficiently small, and moreover that 
\begin{equation}\label{moverrss}
\sup_{0\leq r\leq r_0}\frac{2\open{m}(r)}{r}<\frac89. 
\end{equation}
This implies in particular that $2M_{\mathrm{in}}/r_0< 8/9$ so that $M>9M_{\mathrm{in}}/8$. 
Let $M_\mathrm{out}:=M-M_{\mathrm{in}}$. 
Let $R_1>r_1$ be such that
\begin{equation}\label{mediumstrip}
   R_1-r_1<\frac{r_1-r_0}{6},
\end{equation}
and define
\[
R_0:=\frac{1}{2}(r_1+R_1).
\]
Let $L_+>0$ and let $W_{*}>0$ be such that
\begin{equation}\label{Lplusc}
|W_{*}|\geq 1+\frac{\sqrt{L_{+}}}{R_{0}}.
\end{equation}
Let $W_{-}>0$ satisfy
\begin{equation}\label{mainc}
|W_{-}|\,e^{\frac{-5M}{2R_{0}(1-\frac{2M}{R_{0}})}}
(1-\frac{2M}{R_{0}})^{3/2}\geq 3 |W_{*}|.
\end{equation}
We remark that since $W_*>0$ and $W_->0$ in this section, the modulus is superfluous but it will be needed below. The same remark 
applies to the time variable which is non-negative in this section but which will be non-positive below and we therefore in some 
situations write the modulus of the time variable, cf. (\ref{supp-as}). We 
are now in a position to specify the initial data.
Let $\open{f}=\open{f}_s+\open{f}_m\,$ be initial data of ADM mass $M$ 
such that 
\[
\supp \open{f}_m \subset [R_{0}, R_1]\times [W_-, \infty[\times [0, L_+]\,,
\]
and 
\begin{equation}\label{checkM}
\int_{R_0}^{R_1}4\pi r^2\open{\rho}_m(r)dr=M_\mathrm{out},
\end{equation}
where $\open{\rho}_m$ is the induced energy density by $\open{f}_m$. 

\textit{Remark 1: }Note that the condition (\ref{checkM}) can be arranged by first 
choosing $h_m$ such that $\supp h_m\subset [R_{0}, R_1]\times [W_-, \infty[\times [0, L_+]$ and then choosing an amplitude $A\in \R_+$ such
that $f_m:=Ah_m$ satisfies (\ref{checkM}).

Before stating the main result in this section we define
\begin{equation}
\kappa_*:=\frac{|W_*|}{\sqrt{1+W_*^2+L_+/R_0^2}}(1-\frac{2M}{R_0})e^{-\frac{M}{R_0(1-\frac{2M}{R_0})}}.
\end{equation}
\begin{theorem}\label{Theorem2}
Assume that $r'_0,r_0,r_1,M_{\mathrm{in}},M, L_+, R_{0}, R_1, W_*, W_{-}$ and $\open{f}$ are given as above,
and consider a solution $f$ of the system (\ref{ee1})-(\ref{ee4}), launched by $\open{f},$ on its maximal existence interval $[0,T[$.
Then $T=\infty$, and 
\begin{equation}\label{supp-as}
   \supp f_m(t)
\subset [R_0+|t\,\kappa_*|, \infty[\times [W_*, \infty[\times [0, L_+],\,
\end{equation}
and the resulting spacetime is future causally geodesically complete.
\end{theorem}

\smallskip
\textit{Remark 2: }The initial data in the theorem is adapted to match the initial 
data constructed in~\cite{AKR2}. However, it is important to note that 
the presence of the steady state given by $f_s$ is not needed if the only aim 
is to construct initial data for proving global existence and geodesic completeness. 
Hence, by taking $f_s=0$ and disregard the parameters $r'_0,r_0$ and $M_{\mathrm{in}}$ 
and thus consider a simpler class of initial, then Theorem 2 can be directly 
compared with the result in~\cite{AKR1}, and it can be seen to be an improvement 
of this result. 

\textit{Proof: }We first notice by following the arguments in~\cite{AKR2}, 
that the only way the matter in the outer region $r\geq R_0$ can affect the static solution is via the metric function $\mu$. 
By dropping the time derivatives in the Vlasov equation
we see that in the remaining equation the factor $e^{\mu-\lambda}$ can be canceled. Hence, the static Einstein-Vlasov system is formulated in terms of
the quantities $f,\,\lambda$ and $\mu_r$ and not $\mu$ itself. Therefore $f,\,\lambda$ and $\mu_r$ remain time independent for $r\leq r_0$. The arguments 
of the proof will therefore mainly concern the outer matter given by $f_m$. 
In particular we will see that the outer matter which initially is moving outwards 
will continue to move outwards, 
and therefore there is no direct interaction with the steady state. 
However, in the last part of the proof which concerns causal geodesic completeness 
the steady state will have an influence. 

Let $[0,t_1[$ be the maximal time interval such that for $t\in [0,t_1[$ and $(r,w,L)\in \supp f_m(t)$, $w>W_*$.
By continuity $t_1>0$. Suppose that $t_1\in ]0,T[$, then we must have $w=W_{*}$ for some $w\in \supp f_m(t_1)$, but we will show that $w>W_{*}$ for all 
$w\in \supp f_m(t_1)$. Thus $t_1=T$ and since the matter stays strictly away 
from $r=0$ it follows that $T=\infty$ in view of~\cite{A2} or~\cite{RRS}. 

Consider a characteristic $(R(s),W(s),L)$ with $R(0)\in [R_0,R_1]$ and let
\[
G(t):=E(R(t),W(t),L)+W(t).
\]
Note that $G>0$. We have by (\ref{char1}) and (\ref{char2}) that
\begin{eqnarray}
\frac{dG(s)}{ds}&=&-\left[\lambda_t(s,R(s))\frac{W(s)}{E(R(s),W(s),L)}+\mu_r(s,R(s)) e^{(\mu-\lambda)(s,R(s))}\right]G(s)\nonumber\\
& &+\frac{Le^{(\mu-\lambda)(s,R(s))}}{R(s)^3E(R(s),W(s),L)}.\label{charG}
\end{eqnarray}
Below we will often suppress the arguments but it should be clear that $R=R(s),\,\mu_r=\mu_r(s,R(s))$ etc.
The quantity $G$, which was first introduced in~\cite{A2}, is not suitable for the purpose here
and the quantity that we will consider is $G(t)e^{\hat{\mu}(t,R(t))}(1-2M/R(t))$. We have
\begin{eqnarray}\label{ddsobject}
\frac{d}{ds}(Ge^{\hat{\mu}}(1-\frac{2M}{R}))&=&-\left[\lambda_t\frac{W}{E}+\mu_r e^{\mu-\lambda}\right]
Ge^{\hat{\mu}}(1-\frac{2M}{R})\nonumber\\
& &+\frac{Le^{\mu-\lambda}}{R^3E}e^{\hat{\mu}}(1-\frac{2M}{R})
+[\hat{\mu}_r\frac{W}{E}e^{\mu-\lambda}+\hat{\mu}_t]Ge^{\hat{\mu}}(1-\frac{2M}{R})\nonumber\\
& &+\frac{2M}{R^2}\frac{W}{E}e^{\mu-\lambda}Ge^{\hat{\mu}}\nonumber\\
&=&-\left[\lambda_t\frac{W}{E}+\check{\mu}_r e^{\mu-\lambda}-\hat{\mu}_t\right]
Ge^{\hat{\mu}}(1-\frac{2M}{R})\nonumber\\
& &+\frac{Le^{\mu-\lambda}}{R^3E}e^{\hat{\mu}}(1-\frac{2M}{R})
-\frac{(1+\frac{L}{R^2})}{E}\frac{m}{R^2}e^{\mu+\lambda}e^{\hat{\mu}}(1-\frac{2M}{R})\nonumber\\
& &+\frac{2M}{R^2}\frac{W}{E}e^{\mu-\lambda}Ge^{\hat{\mu}}.
\end{eqnarray}
Here we used that
\[
\hat{\mu}_r\frac{W}{E}e^{\mu-\lambda}G-\hat{\mu}_r e^{\mu-\lambda}G=-\frac{(1+\frac{L}{R^2})}{E}\hat{\mu}_re^{\mu-\lambda}
=-\frac{(1+\frac{L}{R^2})}{E}\frac{m}{R^2}e^{\mu+\lambda}.
\]
Consider the second last term in (\ref{ddsobject}).
In view of (\ref{e2-lambda}) we have
$$1-\frac{2M}{R(t)}\leq e^{-2\lambda(t,R(t))},$$
and we get
\[
-\frac{(1+\frac{L}{R^2})}{E}\frac{m}{R^2}e^{\mu+\lambda}e^{\hat{\mu}}(1-\frac{2M}{R})
\geq -\frac{(1+\frac{L}{R^2})}{E}\frac{m}{R^2}e^{\mu-\lambda}e^{\hat{\mu}}.
\]
Now, since $W\geq W_{*}$ on $[0,t_1],$ and since $R(t)\geq R_0$, we get in view of (\ref{Lplusc}) that
\begin{eqnarray}
&-&\frac{(1+\frac{L}{R^2})}{E}\frac{m}{R^2}e^{\mu-\lambda}e^{\hat{\mu}}
+\frac{2M}{R^2}\frac{W}{E}e^{\mu-\lambda}Ge^{\hat{\mu}}\nonumber\\
& &\geq \Big(-\frac{(1+\frac{L}{R^2})}{E}
+\frac{2WG}{E}\Big)\frac{m}{R^2}e^{\mu-\lambda}e^{\hat{\mu}}\geq 0.
\end{eqnarray}
Hence we have
\begin{equation}\label{ddsclean}
\frac{d}{ds}\Big(Ge^{\hat{\mu}}(1-\frac{2M}{R})\Big)\geq
-\left[\lambda_t\frac{W}{E}+\check{\mu}_r e^{\mu-\lambda}-\hat{\mu}_t\right]
Ge^{\hat{\mu}}(1-\frac{2M}{R}).
\end{equation}
This implies that
\begin{eqnarray}
\displaystyle& &G(t_1)e^{\hat{\mu}(t_1,R(t_1))}(1-\frac{2M}{R(t_1)})\nonumber\\
\displaystyle& &\geq e^{-\int_{0}^{t_1}
\left[\lambda_t(s,R(s))\frac{W}{E}
+\check{\mu}_r(s,R(s))e^{(\mu-\lambda)(s,R(s))}-\hat{\mu}_t(s,R(s))\right]ds}
G(0)e^{\hat{\mu}(0,R(0))}(1-\frac{2M}{R(0)}).\nonumber\\
\label{Gineqmain}
\end{eqnarray}
Let $\gamma$ be the curve
\[
\gamma:=\{(t,r):0\leq t\leq t_1,\, r=R(t)\}.
\]
The time integral in (\ref{Gineqmain}) can be written as
\begin{equation}
\int_{\gamma}e^{(-\mu+\lambda)(t,r)}\lambda_{t}(t,r)\,dr+
\Big(e^{(\mu-\lambda)(t,r)}\check{\mu}_{r}(t,r)-\hat{\mu}_t(t,r)\Big)\,dt.\label{Hcurveweak}
\end{equation}
We will apply Green's formula in the plane to this curve integral.
Let $R_{\infty}\geq R_1+t_1$, so that $f(\cdot,R_{\infty},\cdot)=0$ for $t\in [0,t_1]$. Let $\Gamma$ be the closed curve
\[
\Gamma=\gamma+C_{t_1}+C_{\infty}+C_0,
\]
where
\[
C_{t_1}:=\{(t,r):t=t_1,\;R(t_1)\leq r\leq R_{\infty}\},
\]
\[
C_{0}:=\{(t,r):t=0,\;R(0)\leq r\leq R_{\infty}\},
\]
and
\[
C_{\infty}:=\{(t,r):t_1\geq t\geq 0,\; r=R_{\infty}\}.
\]
We have
\begin{eqnarray}
&\displaystyle\oint_{\Gamma} e^{-\mu+\lambda}\lambda_{t}\,dr+
(e^{\mu-\lambda}\check{\mu}_{r}-\hat{\mu}_t)\,dt&\nonumber\\
&\displaystyle =\int\int_{\Omega}\partial_{t}\left(e^{-\mu+\lambda}
\lambda_{t}\right)-\partial_{r}
\left( e^{\mu-\lambda}\check{\mu}_{r}-\hat{\mu}_t\right)drdt&\nonumber\\
&\displaystyle =\int\int_{\Omega}\partial_{t}\left(e^{-\mu+\lambda}
\lambda_{t}\right)-\partial_{r}
\left( e^{\mu-\lambda}\mu_{r}-\hat{\mu}_t\right)drdt
+\int\int_{\Omega}\partial_{r}
\left(e^{\mu+\lambda}\frac{m}{r^{2}}\right)drdt.&\nonumber\\
\end{eqnarray}
By using (\ref{ee4}) and (\ref{dtmuhat}) this identity can be written
\begin{eqnarray}
& &\phantom{GH} \displaystyle\oint_{\Gamma} e^{-\mu+\lambda}\lambda_{t}\,dr+
(e^{\mu-\lambda}\check{\mu}_{r}-\hat{\mu}_t)\,dt\nonumber\\
& &\displaystyle =\int\int_{\Omega}e^{\mu+\lambda}\left(\frac{2m}{r^{3}}
-4\pi(\rho-p)-8\pi p_T-4\pi e^{2\lambda}j\right)\,drdt\nonumber\\
& &\phantom{G} \displaystyle +\int\int_{\Omega}e^{\mu+\lambda}\left((\mu_r+\lambda_r)\frac{m}{r^2}+4\pi\rho-\frac{2m}{r^3}\right)drdt\nonumber\\
& &\displaystyle=\int\int_{\Omega}4\pi e^{\mu+\lambda}\left[
(\rho+p)e^{2\lambda}\frac{m}{r}+p-2p_T-je^{2\lambda}\right] drdt.\nonumber\\
\end{eqnarray}
Here we used that $\mu_r+\lambda_r=4\pi r(\rho+p)e^{2\lambda}$.
Since $w\geq W_{*}>0$ on $[0,t_1]$ we have in view of (\ref{p}) and (\ref{j}) that $j>p$, and we have 
\[
p(1+e^{2\lambda}\frac{m}{r})=pe^{2\lambda}(1-\frac{2m}{r}+\frac{m}{r})\leq pe^{2\lambda}.
\]
Hence, by dropping the term involving $p_T$ due to sign, we thus obtain
\begin{eqnarray}\label{rectangle}
& &\oint_{\Gamma} e^{-\mu+\lambda}\lambda_{t}\,dr+
(e^{\mu-\lambda}\check{\mu}_{r}-\hat{\mu}_t)\,dt\leq \int\int_{\Omega}4\pi e^{\mu+\lambda}
\rho e^{2\lambda}\frac{m}{r}\, drdt\nonumber\\
& &\leq \int\int_{\Lambda}4\pi e^{\mu+\lambda}
\rho e^{2\lambda}\frac{m}{r}drdt.
\end{eqnarray}
Here $\Lambda=\{(t,r):0\leq t\leq t_1,\; R_{0}\leq r < \infty\},$ and the last inequality
follows since the integrand is nonnegative and $\Omega\subset\Lambda$.
Next we estimate $\rho$ in terms of $j$. Since for $w\in \supp f_m(t),\, 0\leq t\leq t_1$, we have $w\geq W_{*}\geq 1+\sqrt{L_+}/R_0$, we get for $r\geq R_0$, 
\begin{eqnarray}
\rho(t,r)
&\leq&
\frac{\pi}{r^2}\int_{-\infty}^{\infty}\int_0^{\infty} f\,dL\,dw
+ \frac{\pi}{r^2}\int_{-\infty}^{\infty}\int_0^{\infty} w f\,dL\,dw\nonumber\\
&&
{}+
\frac{\pi}{r^2}\int_{-\infty}^{\infty}\int_0^{\infty} \frac{\sqrt{L}}{r} f\,dL\,dw\nonumber\\
&\leq&
3\,\frac{\pi}{r^2}\int_{-\infty}^{\infty}\int_0^{\infty} w f\,dL\,dw =3\,j(t,r).\label{rho3j}
\end{eqnarray}
We estimate the right hand side in (\ref{rectangle}) by using the Vlasov equation from which it follows that
\begin{equation}\label{matteridentity}
\frac{\partial}{\partial t}\left(r^{2}e^{2\lambda}\rho(t,r)\right)=
-\frac{\partial}{\partial r}\left(r^{2}e^{\mu+\lambda}j\right)-re^{\mu+\lambda}2je^{2\lambda}
\frac{m}{r}.
\end{equation}
Since $j(t,R_{0})=0$ and $j(t,R_{\infty})=0$, this gives
\begin{eqnarray*}
\int_{R_{0}}^{R_{\infty}}r^{2}e^{2\lambda(t_1,r)}\rho(t_1,r)dr&-&\int_{R_{0}}^{R_{\infty}}r^{2}e^{2\lambda(0,r)}\rho(0,r)dr\\
&=& -\int\int_{\Lambda}re^{\mu+\lambda}2je^{2\lambda}
\frac{m}{r}drdt.
\end{eqnarray*}
Thus, we get
\begin{equation}\label{Lambdaest}
\int\int_{\Lambda}re^{\mu+\lambda}2je^{2\lambda}
\frac{m}{r}drdt\leq \int_{R_{0}}^{\infty}r^{2}e^{2\lambda}\rho(0,r)dr\leq\frac{1}{4\pi}\frac{M}{1-\frac{2M}{R_{0}}},
\end{equation}
where the last inequality follows in view of (\ref{e2-lambda}) and (\ref{adm}).
Using inequality (\ref{rho3j}) we therefore have
\begin{eqnarray}
4\pi\int\int_{\Lambda}e^{\mu+\lambda}\rho e^{2\lambda}
\frac{m}{r}drdt&\leq& 4\pi\int\int_{\Lambda}e^{\mu+\lambda}3je^{2\lambda}
\frac{m}{r}drdt\nonumber\\
&\leq& 4\pi\frac{1}{R_{0}} \int\int_{\Lambda}re^{\mu+\lambda}3je^{2\lambda}
\frac{m}{r}drdt\nonumber\\
&\leq&\frac32\frac{M}{R_{0}(1-\frac{2M}{R_{0}})}.\label{domainest}
\end{eqnarray}
In order to obtain an estimate for (\ref{Hcurveweak}) it remains to estimate the boundary terms since
\[
\int_{\gamma}...=\oint_{\Gamma}...\,-\,\int_{C_{t_1}}...\,-\,\int_{C_{\infty}}...\,-\,\int_{C_0}...
\]
First we notice that the curve integral along $C_{\infty}$ vanishes since both $p$ and $j$, which determine $\check{\mu}_r$ and $\hat{\mu}_t$, vanishes
for $r=R_{\infty}$. 
Since $j\geq 0$ we have that $\lambda_t\leq 0$, which implies that the integral along $C_0$ can be dropped due to sign noticing the orientation of $C_0$.
The term along $C_{t_1}$ can be estimated as follows.
\begin{eqnarray}
& &\Big|\int_{C_{t_1}}e^{(-\mu+\lambda)(t,r)}\lambda_{t}(t,r)dr+
\Big(e^{(\mu-\lambda)(t,r)}\check{\mu}_{r}(t,r)-\hat{\mu}_t(t,r)\Big)dt\Big|\nonumber\\
& &=\Big|\int_{R(t_1)}^{R_{\infty}}e^{(-\mu+\lambda)(t_1,r)}\lambda_{t}(t_1,r)\,dr\Big|\leq
\int_{R_0}^{\infty}4\pi r\,e^{2\lambda}|j(t_1,r)|\,dr\nonumber\\
& &\leq \frac{M}{R_{0}(1-\frac{2M}{R_{0}})}.\label{jbdryest}
\end{eqnarray}
In the first inequality we used (\ref{ee3}) and in the second we used (\ref{e2-lambda}), the fact that $|j|\leq\rho$, and (\ref{adm}).
We have thus obtained the estimate
\[
\int_{\gamma} e^{-\mu+\lambda}\lambda_{t}\,dr+
(e^{\mu-\lambda}\check{\mu}_{r}-\hat{\mu}_t)\,ds\leq \frac{5M}{2R_{0}(1-\frac{2M}{R_{0}})}.
\]
Inserting this into the main inequality we get
\begin{equation}\label{ineqt1}
G(t_1)e^{\hat{\mu}(t_1,R(t_1))}(1-\frac{2M}{R(t_1)})
\geq e^{\frac{-5M}{2R_{0}(1-\frac{2M}{R_{0}})}}
G(0)e^{\hat{\mu}(0,R(0))}(1-\frac{2M}{R(0)}).\nonumber
\end{equation}
Noticing that $\hat{\mu}$ is monotone in $r$ and nonpositive, and that $R(0)\geq R_0$, we obtain the inequality
\begin{eqnarray}
G(t_1)
&\geq& e^{\frac{-5M}{2R_{0}(1-\frac{2M}{R_{0}})}}
G(0)e^{\hat{\mu}(0,R_{0})}(1-\frac{2M}{R_{0}})\nonumber\\
&\geq& e^{\frac{-5M}{2R_{0}(1-\frac{2M}{R_{0}})}}
G(0)\sqrt{\frac{R_{0}-2M}{R_{0}}}(1-\frac{2M}{R_{0}}).\label{ineqGt1}
\end{eqnarray}
Here we made use of the estimate
\begin{equation}\label{hatmuest}
   \hat{\mu}(t,R_0)
   \geq -\int_{R_0}^{\infty}\frac{M\,d\eta}{\eta^2(1-\frac{2M}{\eta})}=\frac12\log{\Big(1-\frac{2M}{R_0}\Big)}.
\end{equation}
We have that $G(0)>W(0)\geq W_-$, and in view of (\ref{Lplusc}) we also have $3W(t)\geq G(t)$ on $[0,t_1]$. We now use the condition (\ref{mainc}) and obtain
\begin{equation}
3W(t_1)\geq G(t_1)> 3W_{*}.\nonumber
\end{equation}
Thus $W(t_1)>W_{*}$, and necessarily we have $t_1=T$. As was pointed out in the beginning of the proof, since matter stay strictly away from the centre of symmetry, $T=\infty$, cf.~\cite{A2} or~\cite{RRS}. Let us next show that (\ref{supp-as}) holds. From 
the characteristic equation (\ref{char1}) we can conclude that $R(t)\geq R_0+|\kappa_* t|$ in view of the estimates
\[
e^{-\lambda(t,R(t))}=\Big(1-\frac{2m(t,R(t))}{R(t)}\Big)^{1/2}\geq \Big(1-\frac{2M}{R_0}\Big)^{1/2},
\]
and $\mu(t,R(t))=\hat{\mu}(t,R(t))+\check{\mu}(t,R(t))\geq\hat{\mu}(t,R_0)+\check{\mu}(t,R_0)$ where $\hat{\mu}(t,R_0)$ is estimated by (\ref{hatmuest})
and
\[
\check{\mu}(t,R_0)=-\int_{R_0}^{\infty}4\pi\eta p e^{2\lambda(t,\eta)}\,d\eta \geq -\frac{M}{R_0(1-\frac{2M}{R_0})}.
\]
The latter inequality is analogous to (\ref{jbdryest}). We remark that the estimate $R(t)\geq R_0+|\kappa_* t|$ is rough and can be improved by using
arguments from~\cite{AR1}.
In order to complete the proof of the theorem we have to show that any causal geodesic is complete.
We follow the argument in~\cite{AKR1} and introduce the coordinates
\[
x^0 = t,\ x^1 = r\sin\theta\cos\phi,\ x^2 = r\sin\theta\sin\phi,\
x^3 = r \cos\theta.
\]
In these coordinates the metric becomes
\[
g_{00} = -e^{2\mu},\ g_{0a} = 0,\
g_{ab} = \delta_{ab} + (e^{2\lambda}-1)\frac{x_a x_b}{r^2},
\]
where Latin indices $a,b$ run from $1$ to $3$ and $x_a = \delta_{ab}x^b$.
Let us now consider
an arbitrary future directed, time-like or null geodesic,
i.e., a solution $(x^\gamma (\tau),p^\gamma(\tau))$ of the geodesic equations
\begin{eqnarray*}
   \frac{dx^{\gamma}}{d\tau} & = & p^{\gamma}, \\
   \frac{dp^{\gamma}}{d\tau} & = & - \Gamma^{\gamma}_{\delta\epsilon}\,p^{\delta}p^\epsilon,
\end{eqnarray*}
where Greek indices $\gamma,\delta,\epsilon$ run from $0$ to $3$,
$\Gamma^{\gamma}_{\delta\epsilon}$ are the Christoffel symbols,
and
\[
p^0 >0,\ g_{\gamma\delta} p^\gamma p^\delta = - m^2 \leq 0.
\]
We notice that before we had $m=1$ which means that the particles in our system 
have rest mass $1$, but
for causal geodesic completeness we need to consider any $m\geq 0$.
Such a geodesic exists on a maximally
extended interval $[0,\tau_+[$,
and future geodesic completeness means that $\tau_+=\infty$
for all such geodesics.

The following relations between the variables $r$, $w$, $L$, and $p^\gamma$ hold:
\begin{eqnarray*}
E
&=& e^{\mu} \, p^0,\\
w
&=& \frac{x_a p^a}{r}\,e^{\lambda},\\
\frac{L}{r^2}
&=&
\delta_{ab} p^a p^b - \left(\frac{x_a p^a}{r}\right)^2,
\end{eqnarray*}
where we now re-define
\[
   E=E(r,w,L)=\sqrt{m^2+w^{2}+L/r^{2}}.
\]
Since $dt/d\tau = p^0 > 0$ we can re-parameterize the geodesic
by coordinate time $t\in [0,t_+[$. We remark that since there is no matter 
in the region $r\leq r'_0$ the arguments in~\cite{A2,RRS}, which apply to 
any causal geodesic, imply that $t_+=\infty$. This will nevertheless be shown below. 
The arguments in~\cite{A2,RRS} are however not sufficient to conclude that 
$\tau_+=\infty$ since $E$ and $p^0$ may grow in time. 
We now show that $E$ and $p^0$, in the present situation, are bounded for any 
causal geodesic which then implies that $\tau_+=\infty$. 

We call the domains $[0,r'_0]$, $[r'_0,R_0]$, 
and $[R_0,\infty[$ the inner vacuum region, the steady state region, 
and the outer matter region respectively. 

Our strategy is to first show that a geodesic with sufficiently large $E$, 
for which $R\leq R_0$, will at a later time travel outwards for all times. 
We will then apply results from~\cite{A2}. 

\textit{Remark 3: }The analysis given below rests in part on 
the results in the previous part of the proof that rapidly outgoing 
characteristics will continue to move outwards. However, since we have good control 
of the metric and the matter in the domain $R(t)\leq R_0+\kappa_* t$ there are 
alternative approaches to the one given below to obtain a bound on $E$. 

Consider a causal geodesic 
$(R(t),W(t),L)$. We now re-define the quantities $W_*, W_-$ and $\kappa_*$ to adapt 
these to the causal geodesic we now consider. 
Let 
\[
W^g_*:=\max\{1+\frac{\sqrt{L_{+}}}{R_{0}},m+\frac{\sqrt{L}}{{r'}_{0}}\},
\]
and let $W^g_->0$ satisfy
\[
W^g_-\,e^{\frac{-5M}{2R_{0}(1-\frac{2M}{R_{0}})}}
(1-\frac{2M}{R_{0}})^{3/2}\geq 3 |W^g_{*}|.
\]
Define
\[
\kappa^g_*:=\frac{|W^g_*|}{\sqrt{1+{W^g_*}^2+L/{r'}_0^2}}(1-\frac{2M}{R_0})e^{-\frac{M}{R_0(1-\frac{2M}{R_0})}}.
\] 
We have that along a geodesic 
\begin{equation}\label{Etdot}
\frac{dE}{dt}=-\Big[\lambda_t\frac{W}{E}+\mu_r e^{\mu-\lambda}\Big]W,
\end{equation}
\begin{equation}\label{Wtdot}
\frac{dW}{dt}=-\Big[\lambda_t W+\mu_r e^{\mu-\lambda}E-e^{\mu-\lambda}\frac{L}{r^3E}\Big],
\end{equation}
and 
\begin{equation}\label{Rtdot}
\frac{dR}{dt}=\frac{W}{E}e^{(\mu-\lambda)(t,R(t))}.
\end{equation}
To control solutions of equation (\ref{Rtdot}) we need to estimate $e^{\mu-\lambda}$. 
We have in view of (\ref{moverrss}), 
\begin{eqnarray}\label{hatmuest2}
   \hat{\mu}(t,r'_0)&\geq&-\int_{r'_0}^{r_0}\frac{m(t,r)\,dr}{r^2(1-\frac{2m(t,r)}{r})}
-\int_{r_0}^{R_0}\frac{M_{\mathrm{in}}\,dr}{r^2(1-\frac{2M_{\mathrm{in}}}{r})}-\int_{R_0}^{\infty}\frac{M\,dr}{r^2(1-\frac{2M}{r})}\nonumber\\
&\geq&-4\log{\frac{r_0}{r'_0}}+\frac12\log{\Big(1-\frac{2M_{\mathrm{in}}}{r_0}\Big)}+\frac12\log{\Big(1-\frac{2M}{R_0}\Big)}.\nonumber 
\end{eqnarray}
Moreover, since the steady state is a given regular solution we have that $p$ 
is uniformly bounded on $r\in [r'_0,r_0]$. This gives 
\begin{eqnarray}
\check{\mu}(t,r)&=&-\int_{r'_0}^{r_0}4\pi r\,e^{2\lambda}p\,dr
-\int_{r_0}^{R_{0}}4\pi r\,e^{2\lambda}p\,dr-\int_{R_{0}}^{\infty}4\pi r\,e^{2\lambda}p\,dr\nonumber
\\
&\geq&-C-\frac{M_{\mathrm{in}}}{r_{0}(1-\frac{2M_{\mathrm{in}}}{r_{0}})}-\frac{M}{R_{0}(1-\frac{2M}{R_{0}})},\label{jbdryest2}
\end{eqnarray}
in analogy with estimate (\ref{jbdryest}) and where (\ref{moverrss}) was used to 
bound $e^{2\lambda}$ on $[r'_0,r_0]$. 
Since $e^{-\lambda}=\sqrt{1-2m(t,r)/r}$ is strictly positive we have that for 
some constant $C_0>0$ 
\begin{equation}\label{eml}
e^{(\mu-\lambda)(t,r)}\geq C_0 \mbox{ for }r\in [0,\infty[. 
\end{equation}
Define 
\[
P:=\frac{C_0}{\sqrt{m^2+1+L/{r'}_0^2}}. 
\]
It follows that a geodesic $(R,W,L)$ with $|W|\geq 1$ and $R\geq r'_0$ satisfies 
\begin{equation}\label{speedo}
\Big|\frac{dR}{dt}\Big|\geq P. 
\end{equation}
Since the aim is to show that 
$E$ is bounded we can assume that for some $t_0\in [0,t_+[$ 
\begin{equation}\label{t0}
E(t_0)\geq Y_1e^{Y_2}+m+\frac{\sqrt{L}}{r'_0}, 
\end{equation}
where $Y_1\geq W^g_-$ and where $Y_2\geq 0$ will be specified below. Note 
that this condition implies that 
\[
|W(t_0)|\geq Y_1e^{Y_2}.
\] 
We will consider the cases $R(t_0)\geq R_0$; $r'_0\leq R(t_0)\leq R_0$; 
and $R(t_0)<r'_0$, and the subcases that $W(t_0)$ is 
positive or negative. Consider the first case with $W(t_0)$ positive. 
In this case we can directly refer to the arguments in the previous part of 
the proof, using the adapted quantities introduced above, noticing that 
$G(t_0)\geq E(t_0)\geq W^g_-$, to conclude that 
the geodesic will travel outwards for all times with $R(t)\geq R_0+\kappa^g_*\,t$. 
If instead $W(t_0)$ is negative we can assume without loss of generality 
that the geodesic will have $R(t)=R_0$ 
at some time $t$. This follows in view of (\ref{speedo}) since the geodesic 
must otherwise have $|W(t)|\leq 1$ at some time $t$ but then $E(t)$ is 
bounded by $\sqrt{m^2+1+L/R_0^2}$ and the argument can be restarted 
at some later time $t_0$ for which (\ref{t0}) holds. Hence we are 
in the situation of the second case which we will treat below. 

Before continuing with the remaining 
cases we need some auxiliary results. 
The right hand side of (\ref{Wtdot}) depends locally on $r$ except for 
the function $\mu$. Now $\mu\leq 0$ and since the steady state situated in 
$r'_0\leq r\leq r_0$ is a given regular static solution, it follows that 
there is a constant $\Gamma>0$, such that for $|W|\geq 1$ we have 
\begin{equation}\label{GammaW}
\Big|\frac{dW}{dt}\Big|\leq\Gamma|W|,\;\mbox{for }r\in [r'_0,R_0]. 
\end{equation}
Let 
\[
q:=\frac{\Gamma(R_0-r'_0)}{P}.
\]
We choose $Y_2$ such that $Y_2\geq 2q+1.$ 
We return to the remaining cases. We note that in view of (\ref{Etdot}), $E$ 
is constant in the inner vacuum region. Hence, in the case 
where $R(t_0)\leq r'_0$, 
we can without loss of generality assume that $W(t_0)$ is positive and $R(t_0)=r'_0$, 
since if $W<0$ the geodesic will continue inwards for some time, with 
constant $E$, reach a turning point where $W=0$ if $L>0$, 
and then it will travel outwards, with positive $W$, and it cannot 
change sign until it hits the steady state region. We remark that 
if $L=0$ the geodesic will pass through the origin and the ``turning point'' 
will be $r=0$ where the sign of $W$ changes. 
Assume now that $R(t)\leq R_0$ on the time interval $I:=[t_0,t_0+\Delta t]$ 
where $\Delta t=(R_0-r'_0)/P$. We then have in view of (\ref{Wtdot}) that 
\[
\log{W(t)}\geq \log{Y_1}+Y_2-\Gamma\Delta t\geq \log{Y_1}+q+1, 
\] 
for $t\in I$. This implies in particular, since $W(t)\geq 1$, that 
\[
\frac{|W(t)|\,e^{\mu-\lambda}}{\sqrt{m^2+W(t)^2+L/R(t)^2}}\geq P
\]
on $I$. In view of (\ref{Rtdot}) the geodesic has thus necessarily crossed 
the region $r'_0\leq r\leq R_0$ within the time interval $\Delta t$. Hence 
$R(t)=R_0$ with $W(t)\geq W^g_-$ at some $t\in I$. 
We can then repeat the arguments in the first case to conclude that 
the geodesic continues to travel outwards. 
We turn to the case where $R(t_0)\in [r'_0,R_0]$. In the case that $W(t_0)$ is 
positive the argument from the previous case applies. If $W(t_0)$ is negative 
we have by the same argument that on the time interval $I:=[t_0,t_0+\Delta t]$ 
the maximum change for $\log{|W|}$ is $q$ so that $R(t)=r'_0$ at some $t$ with 
$\log{|W(t)|}\geq \log{Y_1}+q+1$. 
Thus we are back in 
the case where the geodesic is in the interior vacuum region and the geodesic 
will turn and have $R(t)=r'_0$ at some later time $t$ with reversed sign on $W(t)$ 
so that $\log{W(t)}\geq Y_1+q+1$ and we can repeat the previous argument. 

To conclude we have reduced the situation so that we only need to consider geodesics 
that satisfy $R(t)\geq R_0+\kappa^g_* (t-t_2)$ for $t\geq t_2$ for some $t_2\geq 0$. 
In the case of a timelike geodesic associated to a particle 
upper bounds on $G=E+W$ and $H:=E-W$, and thus on $E$, are 
obtained in~\cite{A2} under the assumption that there is no matter in 
the domain $\{(t,r):r\leq \epsilon\}$. These bounds are however time dependent. 
The arguments in~\cite{A2} do not depend on the rest mass $m$ and are thus unchanged 
for a causal geodesic. If we apply these arguments in the present situation, where 
the geodesic satisfies $R(t)\geq R_0+\kappa^g_* (t-t_2)$, 
the bounds (4.18) and (4.19) in~\cite{A2} can in fact be seen to be time independent.
Indeed, we need to reconsider the estimates in~\cite{A2} leading to these bounds.
For simplicity we put $t_2=0$. 
Let us here only consider the bound for $G$ since the bound for $H$ follows analogously. We thus reconsider the estimates of (4.14), (4.15) and (4.17) in~\cite{A2}.
For the inequality (4.14) in~\cite{A2} we now get
\begin{equation}
\big|\int_{C_t}4\pi rje^{2\lambda}dr\big|\leq \frac{1}{R_0+|\kappa^g_* t|}\frac{1}{\Big(1-\frac{2M}{R_0+|\kappa^g_* t|}\Big)}\int_0^{\infty}
4\pi r^2\rho\, dr\leq C(M,R_0).
\end{equation}
Since
\[
\frac{L}{E}\leq R\sqrt{L},
\]
the bound (4.15) in~\cite{A2} is replaced by
\begin{equation}
\frac{Le^{\mu-\lambda}}{R^3E}\leq \frac{\sqrt{L}}{(R_0+|\kappa^g_* t|)^2}. 
\end{equation}
Finally, for the term (4.17) in~\cite{A2} we get
\begin{equation}
\int\int_{\Omega} 8\pi(\rho-p)e^{\mu+\lambda}\,dtdr\leq\int_0^t\int_{R_0+|\kappa^g_* s|}^{\infty} \frac{8\pi r^2\rho}{(R_0+|\kappa^g_* s|)^2}\,drds\leq C(M,R_0).
\end{equation}
These estimates turn (4.18) in~\cite{A2} into the time independent bound
\[
G(t)\leq C(M,R_0).
\]
By an analogous argument it follows that also $H(t)\leq C(M,R_0)$, and therefore $E(t)\leq C(M,R_0)$.
This shows that $E$ is bounded for any causal geodesic. 
Now, since $\mu$ is bounded from below it follows that $p^0$ and $p^a$ are bounded 
and thus $t_+=\infty$, and since $dt/d\tau=p^0$ we have that $\tau_+=\infty$, 
and the proof of the theorem is complete.
\begin{flushright}
$\Box$
\end{flushright}
For the proof of Corollary~\ref{Corollary1} a slightly different set of initial data 
is needed. The reason is that there is no result in the literature which says 
that there are steady states of the Einstein-Vlasov system for which $m/r$ is 
arbitrarily small everywhere. By numerical simulations,using the code developed in 
the work~\cite{AR3}, we find nevertheless evidence that this is true. 
It is on the other hand known that there are steady states for which $2m/r$ can be 
arbitrarily close to $8/9$, cf.~\cite{A4} and~\cite{A3}. 

The set up below is similar to our previous set up with the difference that 
the inner matter which is supported in $[r'_0,r_0]$ is not given by a steady state. 

Let $0<r'_0<r_0<r_1$ be given and put $M=r_1/2$. Let 
$L_+>0$ and let $M_{\mathrm{out}}<M$ be such that  
\begin{equation}\label{compactin}
\frac{M-M_{\mathrm{out}}}{r'_0}<\frac12.
\end{equation}
Put $M_{\mathrm{in}}:=M-M_{\mathrm{out}}$. 
Let $R_1>r_1$ be such that
\begin{equation}\label{mediumstrip}
   R_1-r_1<\frac{r_1-r_0}{6},
\end{equation}
and define
\[
R_0:=\frac{1}{2}(r_1+R_1).
\]
Let $W_{*}>0$ satisfy 
\begin{equation}\label{Lplusc2}
|W_{*}|\geq 1+\frac{\sqrt{L_{+}}}{r'_{0}}. 
\end{equation}
Let $\open{f}=\open{f}_{\mathrm{in}}+\open{f}_{\mathrm{out}}$ be such that 
\begin{equation}\label{cn1}
\supp{\open{f}_{\mathrm{in}}}\subset [r'_0,r_0]\times ]W_{1},\infty]\times [0,L_+],
\end{equation}
\begin{equation}\label{cn2}
\supp{\open{f}_{\mathrm{out}}}\subset [R_0,R_1]\times ]W_2,\infty]\times [0,L_+],
\end{equation}
where $W_1>0$ and $W_2>0$ satisfy 
\begin{equation}\label{mainc1}
|W_{1}|\,e^{-\Big(\frac{5M_{\mathrm{in}}}{2r'_{0}(1-\frac{2M_{\mathrm{in}}}{r'_{0}})}+\frac{5M}{2R_0(1-\frac{2M}{R_0})}\Big)}
(1-\frac{2M_{\mathrm{in}}}{r'_{0}})^{3/2}(1-\frac{2M}{R_0})^{1/2}\geq 3 |W_{2}|,
\end{equation}
\begin{equation}\label{mainc2}
|W_{2}|\,e^{-\Big(\frac{3M_{\mathrm{in}}}{2r'_0(1-\frac{2M_{\mathrm{in}}}{r'_0})}+\frac{-5M}{2R_{0}(1-\frac{2M}{R_{0}})}\Big)}
(1-\frac{2M}{R_{0}})^{3/2}\geq 3 |W_{*}|,
\end{equation}
and such that 
\begin{equation}\label{MinMout}
M_{\mathrm{in}}=\int_{r'_0}^{r_0}4\pi r^2\open{\rho}_{\mathrm{in}}\,dr,\;\; 
M_{\mathrm{out}}=\int_{R_0}^{R_1}4\pi r^2\open{\rho}_{\mathrm{out}}\,dr.
\end{equation}
Define 
\[
\kappa'_{*}=\frac{|W_*|}{\sqrt{1+W_*^2+L_+/{r'}_0^2}}A^{3/2}B,
\]
where 
\[
A:=\min\{1-\frac{2M_{\mathrm{in}}}{r_0'},1-\frac{2M}{R_0}\},
\]
and 
\[
B:=e^{-\Big(\frac{M_{\mathrm{in}}}{r'_0(1-\frac{2M_{\mathrm{in}}}{r'_0})}+
\frac{M}{R_0(1-\frac{2M}{R_0})}\Big)}.
\]
\begin{theorem}\label{Theorem3}
Assume that $r_0', r_0, r_1, R_{0}, R_1, L_+, M_{\mathrm{out}}, W_*, W_{1}, W_{2}$ and $\open{f}$ are given as above,
and consider a solution $f$ of the system (\ref{ee1})-(\ref{ee4}), launched by $\open{f},$ on its maximal existence interval $[0,T[$.
Then $T=\infty$, and 
\begin{equation}\label{supp-as2}
   \supp f(t)\subset [r'_0+|t\,\kappa'_*|, \infty[\times 
[W_*, \infty[\times [0, L_+],\,
\end{equation}
and the resulting spacetime is future causally geodesically complete.
\end{theorem}
\textit{Proof: }
The proof follows to a large extent the previous proof and the set up is identical. 
For a characteristic originating from $[R_0,R_1]$, corresponding to the outer matter, 
we consider the same quantity 
\begin{equation}\label{originalquantity}
G(t)e^{\hat{\mu}(t,R(t))}(1-\frac{2M}{R(t)})
\end{equation}
as above. We follow the steps identically until we reach inequality (\ref{rectangle}) 
where we replace $\Lambda$ 
by $\Lambda':=\{(t,r):0\leq t\leq t_1, r'_0\leq r\leq \infty\}$. 
The reason for this modification is that there is no matter at $r=r'_0$ and this fact 
guarantees that the boundary terms which result from the identity 
(\ref{matteridentity}) are zero as before. 
The estimate (\ref{Lambdaest}) is now replaced by
\begin{eqnarray}
\int\int_{\Lambda'}re^{\mu+\lambda}2je^{2\lambda}\frac{m}{r}\,drdt &\leq&\int_{r'_0}^{\infty}r^2e^{2\lambda}\rho(0,r)\, dr\nonumber\\
& &\leq \int_{r'_0}^{R_0}r^2e^{2\lambda}\rho(0,r)\, dr+\int_{R_0}^{\infty}r^2e^{2\lambda}\rho(0,r)\, dr\nonumber\\
& & \leq \frac{1}{4\pi}\frac{M_\mathrm{in}}{1-\frac{2M_{\mathrm{in}}}{r'_0}}+\frac{1}{4\pi}\frac{M}{1-\frac{2M}{R_0}}.\label{primedomainest}
\end{eqnarray}
Using this estimte in (\ref{domainest}) we get 
\begin{eqnarray}
4\pi\int\int_{\Lambda}e^{\mu+\lambda}\rho e^{2\lambda}
\frac{m}{r}drdt&\leq& 4\pi\int\int_{\Lambda}e^{\mu+\lambda}3je^{2\lambda}
\frac{m}{r}drdt\\
&\leq& 4\pi\frac{1}{R_{0}} \int\int_{\Lambda}re^{\mu+\lambda}3je^{2\lambda}
\frac{m}{r}drdt\nonumber\\
&\leq&\frac32\frac{M_{\mathrm{in}}}{r'_{0}(1-\frac{2M_{\mathrm{in}}}{r'_{0}})}+\frac32\frac{M}{R_{0}(1-\frac{2M}{R_{0}})},\label{set}
\end{eqnarray}
in place of (\ref{domainest}).
The remaining estimates are unchanged and the condition (\ref{mainc2}) then ensures 
that the characteristic we consider satisfies $|W(t_1)|>|W_*|$. 

For a characteristic originating from the interval $[r'_0,r_0]$ we define the time $t_{R_0}$ such that if the characteristic reaches $r=R_0$ in the time interval $[0,t_1]$ 
then this happens at $t=t_{R_0}$. Note that on the time interval $[0,t_1]$ 
all particles move outwards so the characteristic can only cross $r=R_0$ once.
On $[0,t_{R_0}]$ we consider the quantity
\begin{equation}\label{innerquantity}
G(t)e^{\hat{\mu}(t,R(t))}(1-\frac{2M_{\mathrm{in}}}{R(t)})
\end{equation}
instead of (\ref{originalquantity}). The influence from the outer matter can only 
enter through the metric coefficient $\mu$ and the actual value of $\mu$ plays no 
role in the proof. We can thus follow the steps in the proof above 
replacing $t_1$ with $t_{R_0}$, $M$ with $M_{\mathrm{in}}$ and $R_0$ with $r'_0$ until we 
reach inequality 
(\ref{rectangle}) where we replace 
$\Lambda$ by $\Lambda':=\{(t,r):0\leq t\leq t_{R_0}, r'_0\leq r\leq \infty\}$. We 
then again use the estimate (\ref{set}) to replace (\ref{domainest}). 
The estimate (\ref{jbdryest}) 
of the boundary term is slightly changed in this case since the domain of integration is now $[r'_0,\infty]$. 
The estimate (\ref{jbdryest}) now becomes 
\begin{eqnarray}
& &\Big|\int_{C_{t_{R_0}}}e^{(-\mu+\lambda)(t,r)}\lambda_{t}(t,r)dr+
\Big(e^{(\mu-\lambda)(t,r)}\check{\mu}_{r}(t,r)-\hat{\mu}_t(t,r)\Big)dt\Big|\nonumber\\
& &\leq\int_{r'_0}^{R_{0}}4\pi r\,e^{2\lambda}|j(t_{R_0},r)|\,dr+\int_{R_{0}}^{\infty}4\pi r\,e^{2\lambda}|j(t_{R_0},r)|\,dr\nonumber
\\
& &\leq \frac{M_{\mathrm{in}}}{r'_{0}(1-\frac{2M_{\mathrm{in}}}{r'_{0}})}+\frac{M}{R_{0}(1-\frac{2M}{R_{0}})}.\label{jbdryest2}
\end{eqnarray} 
Similarly the estimate (\ref{hatmuest}) 
also gives two terms by the same splitting, cf.~(\ref{hatmuest}). 
Hence (\ref{ineqt1}) becomes 
\begin{eqnarray}\label{ineqtR0}
G(t_{R_0})e^{\hat{\mu}(t_{R_0},R(t_{R_0}))}(1-\frac{2M_{\mathrm{in}}}{R(t_{R_0})})
&\geq& e^{-\Big(\frac{5M_{\mathrm{in}}}{2r'_{0}(1-\frac{2M_{\mathrm{in}}}{r'_{0}})}+\frac{5M}{2R_0(1-\frac{2M}{R_0})}\Big)}\nonumber\\
& &\times G(0)e^{\hat{\mu}(0,R(0))}(1-\frac{2M_{\mathrm{in}}}{R(0)}),\nonumber
\end{eqnarray}
and inequality (\ref{ineqGt1}) is then replaced by 
\begin{eqnarray}
G(t_{R_0})
&\geq& e^{-\Big(\frac{5M_{\mathrm{in}}}{2r'_{0}(1-\frac{2M_{\mathrm{in}}}{r'_{0}})}+\frac{5M}{2R_0(1-\frac{2M}{R_0})}\Big)}\nonumber\\
& &\times G(0)(1-\frac{2M}{R_0})^{1/2}(1-\frac{2M_{\mathrm{in}}}{r'_{0}})^{3/2}.\nonumber
\end{eqnarray}
Here the estimate (\ref{hatmuest}) was modified by splitting the integration domain 
in the subintervals $[r'_0,R_0]$ and $[R_0,\infty]$ as above which yields 
\begin{equation}\label{hatmuest2}
   \hat{\mu}(0,r'_0)
\geq\frac12\log{\Big(1-\frac{2M_{\mathrm{in}}}{r'_0}\Big)}+\frac12\log{\Big(1-\frac{2M}{R_0}\Big)}.
\end{equation}
We use condition (\ref{mainc1}) and obtain
\[
W(t_{R_0})>W_{2},
\] 
by using the arguments in the previous case. 

On the remaining time interval $]t_{R_0},t_1]$ we again consider the quantity 
$G(t)e^{\hat{\mu}(t,R(t))}(1-2M/R(t))$ as for a characteristic originating from 
$[R_0,R_1]$ and repeat the arguments in that situation. Note here that $R_1$ is in 
this case replaced by $R_1+t_{R_0}$ but this has no influence on the argument. 
Thus the first statement of Theorem~\ref{Theorem3} holds for this class of initial data. We next consider statement (\ref{supp-as2}). 
If we let 
\[
A:=\min\{1-\frac{2M_{\mathrm{in}}}{r'_0},1-\frac{2M}{R_0}\},
\]
we have that $e^{-\lambda(t,r)}\geq A^{1/2}$. We also have 
\begin{equation}\label{muhatinandout}
e^{\hat{\mu}(t,r)}\geq (1-\frac{2M_{\mathrm{in}}}{r'_0})^{1/2}(1-\frac{2M}{R_0})^{1/2},
\end{equation}
in view of inequality (\ref{hatmuest2}). Similarly we get 
\[
e^{\check{\mu}(t,r)} \geq e^{-\Big(\frac{M_{\mathrm{in}}}{r'_0(1-\frac{2M_{\mathrm{in}}}{r'_0})}+\frac{M}{R_0(1-\frac{2M}{R_0})}\Big)}=:B,
\]
analogously to (\ref{jbdryest2}).
Thus if we define
\[
\kappa'_*:=\frac{|W_*|}{\sqrt{1+W_{*}^{2}}+L_+/{r'}_{0}^{2}}A^{3/2}B, 
\]
it follows that (\ref{supp-as2}) holds. 
The proof of causal geodesic completeness follows the steps in the previous proof 
but is easier since in this case there is no steady state present. 
\begin{flushright}
$\Box$
\end{flushright}
\section{Proof of Theorem 1}
\setcounter{equation}{0}
In this section we prove Theorem~\ref{Theorem1}, Corollary~\ref{Corollary1} and Corollary~\ref{Corollary2}.
The results of Theorem~\ref{Theorem2} and Theorem~\ref{Theorem3} are time reversible in the following sense: by taking initial data
as specified in Theorem~\ref{Theorem2} or in Theorem~\ref{Theorem3} but with 
reversed momenta, disregarding the steady state, so that the outgoing particles are 
ingoing, then global existence \textit{to the past} holds and (\ref{supp-as}) 
and (\ref{supp-as2}) become 
\begin{equation}\label{supp-aspast}
   \supp f(t)\subset [a_1+|a_2 t|, \infty[\,\times\, ]-\infty,W_*]\,\times\, [0, L_+],
\end{equation}
where $a_1$ and $a_2$ are equal to $R_0$ and $\kappa_*$ respectively $r'_0$ and 
$\kappa'_*$, 
and spacetime is past causally geodesically complete. 
We denote such initial data by ${\cal I}_s$ and ${\cal I}_r$ respectively. 
We show that there is a sub class of ${\cal I}_s$ and of ${\cal I}_r$ which satisfy 
the conditions in~\cite{AKR2}. These conditions guarantee 
that black holes form to the future, cf. 
also~\cite{AKR3} where an additional argument is given to match the definition of 
a black hole in~\cite{Cu1}. 
We recall the set up and the conditions on the initial data in~\cite{AKR2}.
There are two slightly different initial data sets in~\cite{AKR2} which both guarantee the formation of black holes
and for our purpose any of these will do. Let us here give the details of Case (i) on p.688 in~\cite{AKR2}.

Let $0<r_0<r_1$ be given, put $M=r_1/2$, and fix $0<M_\mathrm{out}<M$ such that
\begin{equation}\label{icnts}
   \frac{2(M-M_\mathrm{out})}{r_0}<\frac{8}{9}.
\end{equation}
Let $R_1>r_1$ be such that
\begin{equation}\label{mediumstrip}
   R_1-r_1<\frac{r_1-r_0}{6},
\end{equation}
and define
\[
R_0:=\frac{1}{2}(r_1+R_1).
\]
Denote by $\open{\rho}\,$ the energy density induced by the initial
distribution function $\open{f}$. It is required that all the matter
in the outer region $[r_0, \infty[$ is initially located in the strip $[R_0,R_1]$,
with $M_\mathrm{out}$ being the corresponding fraction of the ADM mass $M$, i.e.,
\begin{equation}\label{checkM}
\int_{r_0}^{\infty}4\pi r^2\open{\rho}(r)dr = \int_{R_0}^{R_1}4\pi r^2\open{\rho}(r)dr=M_\mathrm{out}.
\end{equation}
Furthermore, the remaining fraction $M_{\mathrm{in}}=M-M_\mathrm{out}$ should be
initially located within the ball of area radius $r_0$, i.e.,
\begin{equation}\label{M-checkM}
\int_{0}^{r_0}4\pi r^2\open{\rho}(r)dr=M_\mathrm{in}.
\end{equation}
If the inner matter is chosen to be a steady state then 
the solution exists for all $r\in [0,\infty[$ and for all times but
generally this is not required and we define the set
\begin{equation} \label{ddef}
   D:=\{(t,r) \in [0,\infty[^2 \mid r \geq \gamma^+(t)\},
\end{equation}
where $\gamma^+$ is an outgoing radial null geodesic originating from $r=r_0>0$, i.e.,
\begin{equation}\label{gamma+}
   \frac{d \gamma^+}{ds}(s)=e^{(\mu-\lambda)(s,\gamma^+(s))},\;\gamma^+(0)=r_0.
\end{equation}
In this set up we use the following expression for the Hawking mass
\[
m(t,r)=M-\int_r^{\infty}4\pi \eta^2\rho(t,\eta)\,d\eta,
\]
instead of (\ref{m}). The analysis in~\cite{AKR2} is restricted to $D$ since 
the dynamics of the inner matter 
is not essential to conclude that a black hole forms. 
We define
\begin{equation} \label{Gammadef}
\Gamma(r_1,R_1) := \sqrt{\frac{R_1-r_1}{R_1+r_1}},
\end{equation}
and we require that the parameter $W_-<0$ satisfies
\begin{equation}\label{Wminuscondition}
\Gamma(r_1,R_1)^2|W_-|^2\geq\frac{10}{d},
\end{equation}
where
\begin{equation}\label{dcond}
d=\min\left\{\frac12,\frac{r_0}{12 R_1},\frac{r_1-r_0}{300 R_1}\right\}.
\end{equation}
We impose the
\smallskip
\noindent
{\bf General support condition:} For all $(r,w,L) \in \supp \fn\,$ the following holds:
\[
r \in ]0,r_0] \cup [R_0,R_1],
\]
and if $r\in [R_0,R_1]$ then $w \leq W_-$ and also
\begin{equation}\label{hypoL}
   0< L\leq \frac{3L}{\eta}\,\open{m}(\eta) +\eta\,\open{m}(\eta),\ \eta\in [r_0,R_1].
\end{equation}
One of the two classes of initial data specified in~\cite{AKR2}, 
which guarantee the formation of black holes, can now be given. Let
\begin{eqnarray}\label{I1def}
{\cal{I}}_B := \Bigl\{ \fn
&\mid&
\fn \ \mbox{is regular, satisfies (\ref{checkM}), (\ref{M-checkM}),
the general support condition,}\nonumber \\
&&
\mbox{and for}\ (r,w,L)\in \supp \fn\ \mbox{with}\ r\in [R_0,R_1],
\sqrt{L}/r_0 \leq 1
\Bigr\}.
\ \label{condL1t1}
\end{eqnarray}
Corollary 2.3 in~\cite{AKR2} shows that if the inner matter is a steady state then 
the solution exist for all Schwarzschild time. We denote the subset of ${\cal I}_B$ 
for which the inner matter is a steady state by ${\cal I}_{B}^s$. 
The initial data sets ${\cal I}_s$ and ${\cal I}_r$ satisfy to a large extent 
the conditions given above and we are now in a position 
to define the subsets ${\cal J}_s\subset {\cal I}_s$ 
and ${\cal J}_r\subset {\cal I}_r$ which satisfy the claims of 
Theorem~\ref{Theorem1}. 
Let 
\begin{equation}\label{L+}
L_+=\min\{r_0^2,r_0 M_{\mathrm{in}}\}. 
\end{equation}
The set ${\cal J}_s$ is defined by 
\begin{equation}\label{Jdef}
{\cal{J}}_s := \Big\{ \fn\in {\cal I}_{s}\mid L_+\; \mbox{satisfies } (\ref{L+})\; \mbox{and } W_-\; \mbox{satisfies}\, (\ref{Wminuscondition})\Big\}.\nonumber
\end{equation}
Note that there is no conflict between the conditions (\ref{mainc}) and 
(\ref{Wminuscondition}) by taking $|W|_-$ sufficiently large. It is clear 
that ${\cal J}_s\subset {\cal I}_{B}^s$. 
Next we consider the initial data set where both the inner and outer matter 
move rapidly inwards. We define 
\begin{equation}\label{J2}
{\cal J}_r:=\Big\{\open{f}=\open{f}_{\mathrm{in}}+\open{f}_{\mathrm{out}}\in {\cal I}_r\mid 
L_+\, \mbox{satisfies } (\ref{L+})\, \mbox{and }W_2\; \mbox{satisfies } (\ref{Wminuscondition})\Big\}.\nonumber 
\end{equation}
Since this set is a subset of ${\cal I}_B$ the proof of the theorem is complete. 
\begin{flushright}
$\Box$
\end{flushright}
\textit{Proof of Corollary 1: }
Let $\epsilon>0$ be given and choose initial data in ${\cal J}_r.$ By evolving the data to the past we have
by (\ref{supp-as2}) that the matter is supported
in the domain $\{(t,r):r\geq r'_0+|\kappa'_* t|\}$. Thus, by evolving to time $t=-T,$ where $T$ is sufficiently large so that
\[
\sup_{r}\frac{m(t,r)}{r}\leq \frac{M}{r'_0+|\kappa'_*|T}\leq \epsilon,
\]
the claim follows by taking as initial data the solution at $t=-T$.
\begin{flushright}
$\Box$
\end{flushright}
\textit{Proof of Corollary 2: }
Consider initial data in ${\cal J}_s$.
The statement then follows by combining Corollary 2.3 in~\cite{AKR2} and 
Theorem~\ref{Theorem2}.
\begin{flushright}
$\Box$
\end{flushright}

\noindent
{\bf Acknowledgment\,:} The author is grateful for discussions with Gerhard Rein and Alan Rendall.

\end{document}